\pdfoutput=1

\documentclass[prl,aps,superscriptaddress,amsmath,amssymb,nofootinbib,twocolumn]{revtex4-2}

\usepackage{braket,verbatim,bm,bbold,amsmath,amssymb,epsfig,float,color}
\usepackage[colorlinks=true,linkcolor=blueG, citecolor=redG, urlcolor=magentaG, bookmarks]{hyperref}
\usepackage{color}
\usepackage[dvipsnames]{xcolor}
\usepackage{enumitem}   
\usepackage{soul}
\usepackage{amsmath,amsfonts,amssymb,stackengine,graphicx}
\setstackgap{S}{-1.5pt}

\newcommand{\nc}{\newcommand}
\nc{\ir}{\mathrm{i}}
\nc{\dd}{\mathrm{d}} 
\nc{\eE}{\mathsf{e}}
\nc{\Tr}{\text{Tr}}
\nc{\id}{\mathbb{I}}
\nc{\I}{\mathcal{I}}
\nc{\F}{\mathcal{F}}
\nc{\M}{\mathcal{M}}
\nc{\Ll}{\mathcal{L}}
\nc{\gap}{\gamma_+}
\nc{\gam}{\gamma_-}

\definecolor{blueG}{RGB}{51, 102, 204}
\definecolor{magentaG}{RGB}{214.2, 40.8, 132.6}
\definecolor{redG}{RGB}{229.5, 51., 102.}

\begin{document}

\title{Smearing of dynamical quantum phase transitions in dissipative free-fermion systems}  

\author{Gilles Parez}
\email{parez@lapth.cnrs.fr}
\affiliation{\it  Laboratoire d'Annecy de Physique Th\'eorique (LAPTh), CNRS,
Universit\'e Savoie Mont Blanc, 74940 Annecy, France}

\author{Vincenzo Alba}
\email{vincenzo.alba@unipi.it}
\affiliation{\it Dipartimento di Fisica, Universit\`a di Pisa, and INFN Sezione di Pisa, Largo Bruno Pontecorvo 3, Pisa, Italy}

\date{\today}

\begin{abstract}
We investigate the Lindblad dynamics of the reduced Loschmidt echo (RLE) in dissipative quadratic fermion systems. Focusing on the case of gain and loss dissipation, we derive general conditions for the persistence of nonanalyticities (so-called dynamical quantum phase transitions) in the time evolution of the RLE. We show that nonanalyticities that are present in the corresponding unitary dynamics can survive under purely gain or purely loss processes, but are completely smeared out as soon as both channels are active, even if one is infinitesimally small. These results hold for generic dissipative Gaussian evolutions, and are illustrated explicitly for the quench from the N\'eel state in the tight-binding chain, as well as for the quantum Ising chain. We also show that the subtle interplay between dissipative and unitary dynamics gives rise to a nested lightcone structure in the dynamics of the RLE, even in cases where this structure is not present in the corresponding unitary evolution, due to coherent cancellations in the phase structure of the wavefunction.
\end{abstract}

\maketitle

 \section{Introduction}Quantum phase transitions are widespread in many-body systems and play a pivotal role in our understanding of quantum matter at equilibrium. They are characterized by nonanalyticities in equilibrium quantities, such as the ground-state energy density, as an external control parameter crosses a critical value. Importantly, equilibrium phase transitions have a nonequilibrium counterpart, the so-called  \textit{dynamical quantum phase transitions} (DQPT)~\cite{heyl2013dynamical,zvyagin2016dynamical,heyl2018dynamical}.  In an isolated quantum system, the simplest protocol to drive the system out of equilibrium is a quantum quench, where the time evolution is generated by a Hamiltonian $H$ acting on an initial state $|\psi_0\rangle$, and the time-dependent state is $|\psi(t)\rangle = \eE^{-\ir t H}|\psi_0\rangle$. In this context, a central quantity in the study of DQPTs is the Loschmidt echo (LE) \cite{peres1984stability}, defined as $ |\langle \psi_0 | \psi(t)\rangle |^2$, and its associated rate function $ \lambda(t) = -\lim_{N\to \infty}1/N\log  |\langle \psi_0 | \psi(t)\rangle |^2$, where~$N$ is the number of degrees of freedom of the system. 
Temporal nonanalyticities of the rate function signal DQPTs, which occur when the time-evolved state becomes orthogonal to the initial one at a critical time $t_c$. The systematic investigation of LE has allowed to detect and probe DQPTs in various nonequilibrium many-body systems, such as critical spin chains and strongly interacting systems \cite{heyl2013dynamical,heyl2014dynamical,andraschko2014dynamical,lacki2019dynamical}, topological insulators \cite{vajna2015topological,schmitt2015dynamical,budich2016dynamical,sedlmayr2019dynamical,zamani2024scaling}, and to investigate DQPTs experimentally in ultra-cold atoms and trapped-ions quantum simulators~\cite{jurcevic2017direct,zhang2017observation,flaschner2018observation,karch2025probing}. 

The LE has a series of limitations. First, it requires the knowledge of the state of the full many-body system, and is typically exponentially small in the system size, rendering it challenging to probe experimentally. Recently, it was demonstrated experimentally \cite{karch2025probing} and analytically ~\cite{halimeh2021local,bandyopadhyay2021observing,parez2025reduced} that DQPTs can be captured by different quasi-local quantities. In particular, we here introduced the reduced Loschmidt echo (RLE) \cite{parez2025reduced}, a quantity closely related to the subsystem LE defined in Ref.~\cite{karch2025probing}, and showed that it is a powerful tool to detect DQPTs. The RLE is the reduced fidelity \cite{parez2022symmetry} between the initial state of a given subsystem~$A$ and its time-evolved version after the quench. Importantly, the RLE reduces to the pure-state LE if both $\rho_A(0)$ and $\rho_A(t)$ are rank-one projectors on pure states, where $\rho_A(t)=\Tr_{\bar A}\rho(t)$ is the reduced density matrix of $A$, and $\bar A$ is its complement, such that $A \cup \bar A$ is the full system.

As a second limitation, the LE is defined as an overlap of pure states, hence pertaining to isolated quantum systems out of equilibrium. However, there are numerous realistic physical contexts where the full quantum system is described by a mixed density matrix $\rho(t)$ instead of a pure state. These include finite-temperature states, or open quantum systems interacting with an environment. Mixed-state DQPT have been considered in those contexts \cite{heyl2017dynamical,bhattacharya2017mixed,lang2018dynamical,sedlmayr2018fate,mera2018dynamical,jafari2024dynamical,nava2024dissipation,nava2025speeding,zhang2025dynamical}, but comparatively much less so than their pure-state analog. A natural question, with direct experimental relevance, is to understand the fate of DQPTs for dissipative dynamics. In particular, if an isolated unitary dynamics admits DQPTs, do these singularities persist in the presence of dissipation?

Here, we solve this problem for open free-fermionic systems subject to gain and loss dissipation, whose time-evolution is determined by the Lindblad master equation \cite{breuer2002theory}. We probe DQPTs for these evolutions with the RLE, and find conditions for their existence based on the dynamics of the related nondissipative model, as well as the gain and loss rates. In particular, we show that if the nondissipative case exhibits DQPTs, then the dissipative evolution with only gain or only loss can preserve the temporal nonanalyticities. However, if both gain and loss are nonzero, even if one is infinitesimally small, the DQPT smears out and the dynamics of the RLE becomes analytic at all times. These results hold for any free-fermion model, and we illustrate them with the dissipative dynamics generated by the tight-binding chain from the N\'eel state, as well as for a dissipative quench in the quantum Ising chain. The RLE is thus a natural tool to probe DQPTs in both isolated and open quantum systems, thereby paving the way for experimental realizations. While finalizing this work, we became aware of related recent results reported in Ref.~\cite{zhang2025dynamical}, where similar observations were obtained using many-body backflow and a mixed-state extension of the full-system Loschmidt echo. 

\section{Quadratic fermion model with dissipation}

We consider an open free-fermion chain interacting with an environment, where fermions can jump in and out of the chain, with rates $\gamma_{\pm}$. This dissipative gain and loss dynamics  is described by the Lindblad master equation~\cite{breuer2002theory}
\begin{subequations}\label{eq:Lind}
\begin{equation}
    \frac{\dd \rho(t)}{\dd t} = \Ll(\rho(t)),
\end{equation}
where $\rho(t)$ is the full-system density matrix, and $\Ll(\rho)$ is the Lindblad operator,
\begin{multline}
   \Ll(\rho)= -\ir [H,\rho] \\+\sum_{j=1}^N\sum_{\alpha ={\pm}}\left(L_{j,\alpha}\rho L^\dagger_{j,\alpha}-\frac 12 \left\{L^\dagger_{j,\alpha}L_{j,\alpha},\rho\right\}\right).
\end{multline}
\end{subequations}
Here, $H$ is a quadratic fermionic Hamiltonian, and the operators $L_{j,\pm}$ model the gain or loss of a fermion on site $j$ during the evolution. We focus on quadratic Lindblad operators, and thus choose $L_{j,+} = \sqrt{\gap}c_j^{\dagger}$ and $L_{j,-} = \sqrt{\gam}c_j$, where $\gamma_\pm \geqslant 0$ are the gain and loss rates, and $c^{(\dagger)}_j$ satisfy the canonical fermionic anticommutation relations $\{c^\dagger_j, c_{j'}\}=\delta_{j,j'}$ and $\{c_j, c_{j'}\}=0$. The fact that the Lindblad operator is quadratic guarantees that the state $\rho(t)$ remains Gaussian throughout the time evolution, provided that the initial state is itself Gaussian. 

\subsection{Reduced Loschmidt echo} 

During the time evolution, we focus on a subsystem~$A$ composed of~$\ell$ contiguous sites, $A=[1,2,\dots,\ell]$. We compare the initial state of $A$ with its time-evolved version using the RLE, defined as~\cite{parez2025reduced}
\begin{equation}
    \F_A(t)  =  \frac{\Tr(\rho_A(0) \rho_A(t))}{\sqrt{\Tr(\rho_A(0)^{2})\Tr(\rho_A(t)^{2})}}. 
\end{equation}
It satisfies $0\leqslant  \F_A(t) \leqslant 1$, with $ \F_A(t)=1$ iff $\rho_A(t)=\rho_A(0)$. Moreover, the RLE reduces to the standard LE when $\rho_A(0)$ and $\rho_A(t)$ project on pure states. A DQPT occurs if there is a time $t_c$ such that the logarithmic RLE becomes singular, or equivalently that the numerator $\Tr(\rho_A(0) \rho_A(t_c))$ approaches zero in the large-$\ell$ limit. Since the numerator is the only nontrivial part which contains signatures of DQPTs \cite{parez2025reduced}, we study the unnormalized RLE, defined simply as $f_A(t) = \Tr(\rho_A(0) \rho_A(t))$, and its logarithmic version $\zeta_A(t) = -1/\ell \log f_A(t)$. 

For Gaussian states which preserve the fermion number, the RLE can be expressed in terms of the two-point correlation matrix 
\begin{equation}
    [C_A(t)]_{x,x'} = \Tr(\rho_A(t)c_x^\dagger c_{x'}), 
\end{equation}
with $x,x'=1,2,\dots,\ell$. For simplicity, we introduce the covariance matrix as $J_A(t)=2C_A(t)-\mathbb{1}_\ell$, and we have~\cite{parez2022symmetry,parez2025reduced} 
\begin{equation}\label{eq:fdet}
    f_A(t) = \det\left(\frac{\mathbb{1}_{\ell}+J_0J_A(t)}{2}\right)
\end{equation}
where $J_0\equiv J_A(0)$. A similar formula exists for Gaussian states which do not preserve the fermion number \cite{parez2025reduced}. 

 \section{DQPTs for dissipative Gaussian evolutions}
 
 We now investigate under what circumstances DQPTs occur in dissipative evolutions generated by a quadratic Lindblad operator from a Gaussian initial state, such that Eq.~\eqref{eq:fdet} holds. The logarithmic RLE reads
 \begin{equation}\label{eq:lft}
\begin{split}
    \zeta_A(t) &= -\frac 1\ell \Tr \log\left( \frac{\mathbb{1}_{\ell}+J_0J_A(t)}{2}\right)\\
    &=-\frac 1\ell\sum_{m=1}^\ell \log\left( \frac{1+\nu_m(t)}{2}\right),
    \end{split}
\end{equation}
where $\nu_m(t)$ are the eigenvalues of the matrix $J_0J_A(t)$. The matrices $J_0,J_A(t)$ are Hermitian but do not commute (except for trivial quench protocols which we do not consider), so that their product is not Hermitian and the eigenvalues can be complex. However, since $\rho_A(0),\rho_A(t)$ are positive semidefinite Hermitian matrices, we have $f_A(t)\geqslant 0$, such that complex eigenvalues $\nu_m$ come in pairs $(\nu_m,\nu_m^*)$. Looking at Eq.~\eqref{eq:lft}, we conclude that if there is a time $t_c$ such that at least one eigenvalue $\nu_M(t)$ approaches $\nu_M(t_c)=-1$ in the large-$\ell$ limit, then there is a DQPT at $t=t_c$, corresponding to a nonanaliticity of the logarithmic RLE. 

To proceed, we relate the condition for the occurrence of a DQPT in the dissipative evolution with the corresponding unitary one. Let us denote by $\tilde C_A(t)$ the correlation matrix for the unitary evolution with the same initial state and Hamiltonian, but where there is no dissipation, i.e., one has $\gamma_\pm=0$. In the following, quantities with a tilde systematically pertain to the corresponding dissipationless evolution. For Gaussian dissipative evolution, we have \cite{alba2021spreading,carollo2022dissipative,murciano2023symmetry} 
\begin{equation}
     C_A(t) = n_\infty(1-b(t))\mathbb{1}_{\ell}+b(t) \tilde C_A(t),
\end{equation}
where we introduced 
\begin{equation}\label{eq:b}
     b(t) = \eE^{-(\gap+\gam)t}, \quad n_\infty=\frac{\gap}{\gap+\gam}.
\end{equation}
 In terms of covariance matrices, this translates to 
\begin{subequations}\label{eq:Jad}
\begin{equation}\label{eq:Jad1}
    J_A(t) = \chi(t) \mathbb{1}_{\ell} + b(t) \tilde J_A(t)
\end{equation}
with 
\begin{equation}\label{eq:chi}
    \chi(t) = (2n_\infty-1)(1-b(t)).
\end{equation}
\end{subequations}
Since $\chi(0)=0$ and $b(0)=1$, the initial covariance matrices are identical, $\tilde J_0 = J_0$. 

The spectrum of the fermionic correlation matrix is bounded between 0 and 1 as a direct consequence of the canonical anticommutation relations of the fermionic operators, which encode the Pauli exclusion principle. As a consequence, the eigenvalues of the dissipationless covariance matrix $\tilde J_A(t)$ are restricted between $-1$ and 1.
In turn, with Eq.~\eqref{eq:Jad}, we find
\begin{equation}
    \lVert J_A(t) \lVert \leqslant \max(|\chi(t) -b(t)|,|\chi(t) +b(t)|)
\end{equation}
and $\lVert J_0 \lVert \leqslant 1$, where $\lVert \bullet \lVert$ is the spectral matrix norm (or 2-norm), which is equal to the largest singular value of the matrix. The modulus of the largest eigenvalue of a square matrix is smaller or equal to its largest singular value, which implies here $ \max_{m} |\nu_m(t)| \leqslant\lVert J_0J_A(t) \lVert $. Using the submultiplicativity of the norm, we finally conclude 
\begin{equation}\label{eq:nuineq-1}
    \max_{m=1,2\dots,\ell} |\nu_m(t)| \leqslant \max(|\chi(t) -b(t)|,|\chi(t) +b(t)|),
\end{equation}
where we recall that $\nu_m(t)$ are the eigenvalues of $J_0J_A(t)$.

From Eqs.~\eqref{eq:b} and \eqref{eq:chi} for $b(t)$ and $\chi(t)$, there are then three possibilities (excluding the unitary dynamics):
\begin{equation}\label{eq:nuineq0}
\max_{m=1,2\dots,\ell} |\nu_m(t)|  
\begin{cases}
< 1 & \text{if} \ \gap>0 \ \text{and} \ \gam>0,\\[6pt]
\leqslant 1 & \text{if} \ \gap>0 \ \text{and} \ \gam=0, \\[6pt]
\leqslant 1 & \text{if} \ \gap=0 \ \text{and} \ \gam>0. \\[6pt]
\end{cases}
\end{equation}

In the first case, which corresponds to a gain and loss process, the inequality is strict. This rules out the possibility of a DQPT during the dissipative dynamics, irrespective of whether the unitary dynamics has a DQPT. When it does, the unitary DQPT is thus smeared out by the presence of gain and loss dissipation. We stress that the conclusion holds even if one or both rate parameters are infinitesimally small.

The two last lines in Eq.~\eqref{eq:nuineq0} correspond to gain-only and loss-only dynamics, respectively. Without loss of generality we focus on the gain-only case, with $\gap>0$, $\gam=0$ and $0 \leqslant \chi(t) =  1-b(t)\leqslant 1$. With Eq.~\eqref{eq:nuineq-1}, this implies $\max_{m} |\nu_m(t)| \leqslant 1$, and hence a DQPT is possible, but not guaranteed. We thus investigate under what conditions DQPTs can occur in the gain-only dynamics.

With Eq.~\eqref{eq:Jad} and the subadditivity of the norm, we~find
\begin{equation}
    \big\lVert J_0 J_A(t) \big\lVert \leqslant (1-b(t)) \big\lVert J_0 \big\lVert + b(t) \big\lVert J_0 \tilde J_A(t) \big\lVert. 
\end{equation}
The spectral matrix norm satisfies $ \max_{m} |\nu_m| \leqslant \big\lVert J_0 J_A \big\lVert$, and hence we are able to
relate the spectra of $J_0J_A(t)$ with its dissipationless analog, 
\begin{equation}\label{eq:nuineq}
     \max_{m=1,2\dots,\ell} |\nu_m(t)| \leqslant 1+b(t)\left(\big\lVert J_0 \tilde J_A(t)\big\lVert-1\right),
\end{equation}
where we used  $\lVert J_0 \lVert \leqslant 1$.
Moreover, in the dissipationless case we have $\max_{m} |\tilde \nu_m(t)|\leqslant \big\lVert J_0 \tilde J_A(t)\big\lVert\leqslant 1$. Therefore, 
if there is a DQPT in the unitary dynamics at time~$\tilde t_c$, i.e., $\max_{m} |\tilde \nu_m(\tilde t_c)|=1$ in the large-$\ell$ limit, then Eq.~\eqref{eq:nuineq} becomes $\max_{m} |\nu_m(\tilde t_c)|\leqslant 1$ and the DQPT can persist in the gain-only dissipative case. If, on the contrary, there is no DQPT in the dissipationless dynamics and the inequality is strict, $\big\lVert J_0 \tilde J_A(t)\big\lVert< 1$, then with Eq.~\eqref{eq:nuineq} we also have $ \max_{m} |\nu_m(t)| < 1$, such that there is no DQPT in the dissipative dynamics. In principle there can also be times $\tilde t_j$ such that $ \big\lVert J_0 \tilde J_A(\tilde t_j)\big\lVert=1$ in the large-$\ell$ limit, but for which there is no DQPT in the unitary dynamics, i.e., no eigenvalues satisfying $\tilde \nu_j(\tilde t_j)=-1$ in the large-$\ell$ limit. In this situation the inequality~\eqref{eq:nuineq} nonetheless reads $\max_{m} |\nu_m(\tilde t_j)| \leqslant 1$, such that we cannot directly rule out the presence of a DQPT in the dissipative dynamics. To do so, we note that the eigenvalues of $J_0J_A(\tilde t_j)$
are continuous functions of the gain rate $\gap$. 
Hence, on the one hand, for $\gap>0$ small enough, all eigenvalues still satisfy $\textrm{Re}(\nu_m(\tilde t_j))>-1$ in the large-$\ell$ limit, such that there is no DQPT. 
On the other hand, for large $\gamma_+$ we have $b(\tilde t_j) \sim 0$ such that $J_A(\tilde t_j) \sim \mathbb{1}_{\ell}$, also implying the absence of DQPT. In fact, the real part of the eigenvalues $\textrm{Re}(\nu_m)$ monotonically increase towards $1$ for increasing~$\gamma_+$. Hence, dissipative dynamics for arbitrary $\gamma_+>0$ cannot exhibit DQPTs at times where the corresponding unitary dynamics is analytical.

We conclude that gain-only (and loss-only) dynamics can exhibit DQPTs only if the corresponding unitary dynamics does, and at the same critical times: $t_c=\tilde t_c$. However, we stress that, while the presence of DQPTs in the unitary dynamics is a necessary condition to observe DQPTs in the dissipative case, it is not a sufficient one. Indeed, there are examples of smooth loss-only dynamics with a corresponding nonanalytic unitary evolution~\cite{sedlmayr2018fate}.

\section{Results for the dissipative N\'eel quench}
As an illustration of our generic results, we focus on the dynamics generated by the tight-binding Hamiltonian 
\begin{equation}
    H=-\frac 12\sum_{j=1}^N(c_{j+1}^\dagger c_{j}+c_{j}^\dagger c_{j+1})
\end{equation}
with periodic boundary conditions. This Hamiltonian preserves the fermion number, $[H,\sum_j c_j^\dagger c_j]=0$, such that Eq.~\eqref{eq:fdet} holds, provided that the initial state is Gaussian.  We work in the large-$N$ limit, where the single-particle energies are $\epsilon_k=\cos k$ for $k\in [-\pi, \pi]$. As initial Gaussian state, we choose the N\'eel state $|\psi_0\rangle = \prod_j c_{2j}^\dagger |0\rangle$, where $|0\rangle$ is the fermionic vacuum, which satisfies $c_j|0\rangle=0$ for all $j$. The dissipationless covariance matrix $\tilde J_A(t)$ has entries \cite{ac2-19}
\begin{equation}\label{eq:JatildeNeel}
    [\tilde J_A(t)]_{x,x'} = (-1)^{x'}\int_{-\pi}^{\pi}\frac{\dd k}{2\pi}\eE^{\ir k(x-x')+2\ir t\cos k}
\end{equation}
with $x,x'=1,2,\dots,\ell$. Importantly, the dissipationless dynamics exhibits DQPTs for times $\tilde t_c^{(m)} = (m+1/2)\pi$, where $m\geqslant 0$ is an integer, signaled by nonanalyticities in the pure-state LE \cite{andraschko2014dynamical} as well as the RLE \cite{parez2025reduced}. 

We now investigate the effect of dissipation on these DQPTs by considering a dissipative evolution generated by a quadratic Lindblad operator with gain and loss.
In practice, we use Eqs.~\eqref{eq:Jad} and \eqref{eq:JatildeNeel} to compute the logarithmic RLE given in Eq.~\eqref{eq:lft}, both numerically and analytically.

\begin{figure}
    \centering
    \includegraphics[width=1\linewidth]{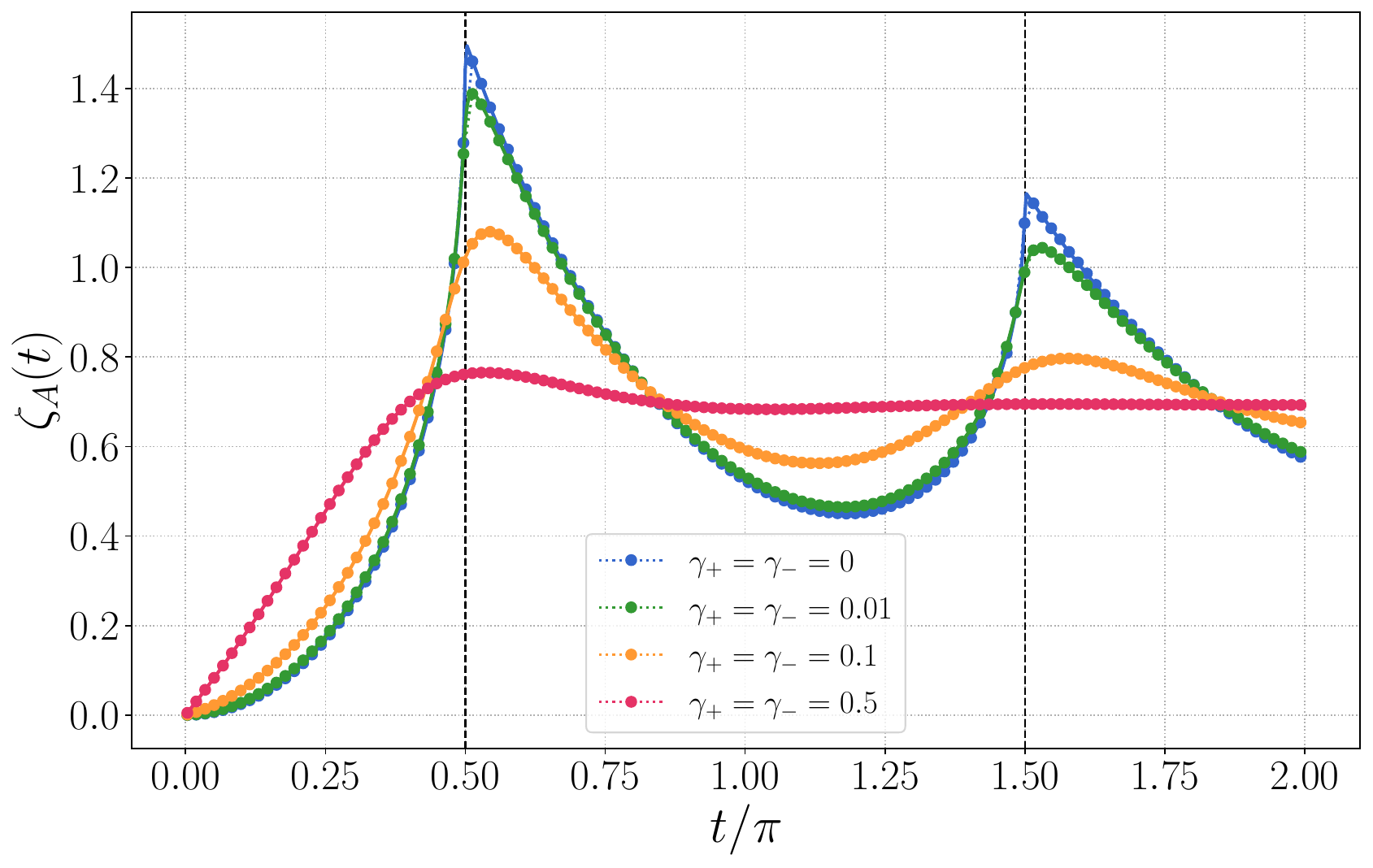}\\
          \includegraphics[width=1\linewidth]{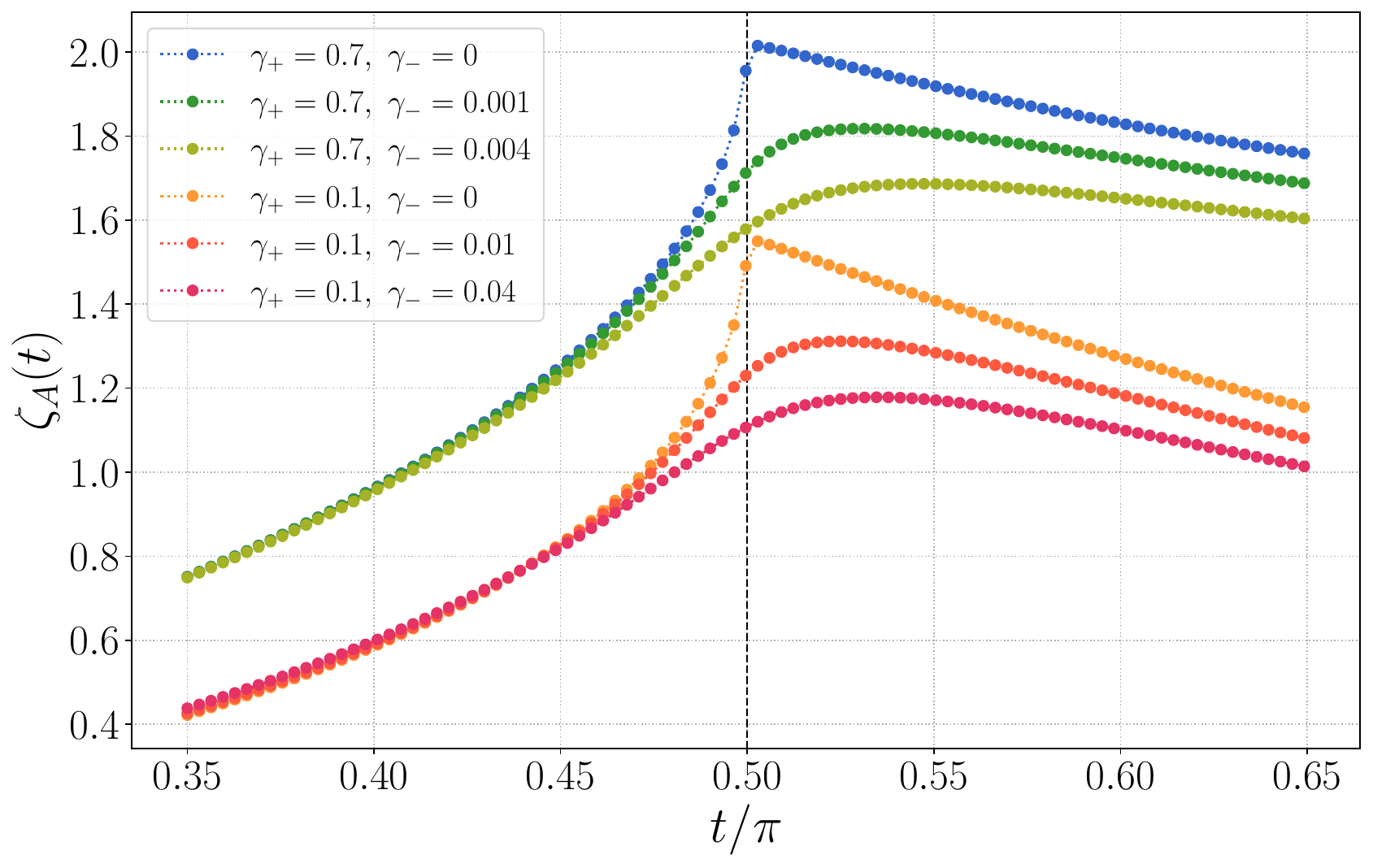}
    \caption{Logarithmic RLE $\zeta_A(t)$ in the dissipative dynamics generated by the tight-binding Hamiltonian from the N\'eel state as a function of $t/\pi$ for $\ell=350$ and various values of gain and loss rates $\gap,\gam$. The symbols are obtained by numerical diagonalization of the covariance matrices in Eq.~\eqref{eq:lft}, and the solid lines in the top panel are the theoretical predictions of Eq.~\eqref{eq:zeta_analy} for the case of equal gain and loss, $\gap=\gam$. Finally, the vertical dashed lines indicate the position of the DQPTs in the dissipationless dynamics. }
    \label{fig:NeelGL}
    \end{figure}

To study analytically the full time evolution of the RLE, we expand it as $\log f_A(t) = \sum_{n=0}^\infty c_n \M_n$, where $c_n$ are the Taylor coefficients of the function $\log((1+x)/2)$ around $x=0$, and $\M_n = \Tr((J_0 J_A(t))^n)$ are the moments of the logarithmic RLE. In the special case of equal gain and loss, i.e., $\gap=\gam$, we can adapt the derivations of Ref.~\cite{parez2025reduced} for the corresponding unitary dynamics. In the sub-hydrodynamic regime where $0 \ll t \ll \ell$, we find
\begin{equation}\label{eq:zeta_analy}
\begin{split}
    \zeta_A(t)= \int_{-\pi}^\pi  \frac{\dd k}{2 \pi} (\log 2 -\eE^{2 t(\ir \cos k-\gamma)})\min(2|v_k| t/\ell,1)&\\ 
    -\int_{-\pi}^\pi  \frac{\dd k}{2 \pi}  \log \left(\frac{1+\eE^{2 t(\ir \cos k-\gamma)}}{2}\right)(1 -\min(2|v_k| t/\ell&,1))
    \end{split}
\end{equation}
where $\gamma=\gap=\gam$ and $v_k=\epsilon_k' =\sin k$. This result can be interpreted in the framework of the quasiparticle picture \cite{calabrese2005evolution,fc-08,alba2017entanglement}, where entangled quasiparticles with opposite velocities $v_k$ and $-v_k$ are emitted from every point in the system and propagate after the quench, spreading entanglement and correlations. We verify the analytical prediction of Eq.~\eqref{eq:zeta_analy} in the top panel of Fig.~\ref{fig:NeelGL} and find a perfect agreement with the numerics for various values of $\gamma$, including the dissipationless case $\gamma=0$. In this case, repeating the argument of Ref.~\cite{parez2025reduced}, the DQPTs are caused by singularities of the function $ \log (1+\eE^{2 \ir t \cos k})$ in Eq.~\eqref{eq:zeta_analy}, and its derivatives. These occur when $\eE^{2 \ir \tilde t_c \cos k}=-1$, but contributions from $k$ and $-k$ cancel in the integral, except for $k=0$. This yields the condition $\eE^{2 \ir \tilde t_c}=-1$, or $\tilde t^{(m)}_c=(m+1/2)\pi$, where $m\geqslant 0$ is an integer. Now, in the case of equal gain and loss dissipation, corresponding to $\gamma>0$, the same argument amounts to searching for singularities of the function $\log (1+\eE^{2 t(\ir \cos k-\gamma)})$. However, since $|\eE^{2 t(\ir \cos k-\gamma)}|<1$ for $t>0$, there are no temporal nonanalyticities, and hence no DQPTs, even for arbitrarily small values of $\gamma$.

\begin{figure}
    \centering
    \includegraphics[width=1\linewidth]{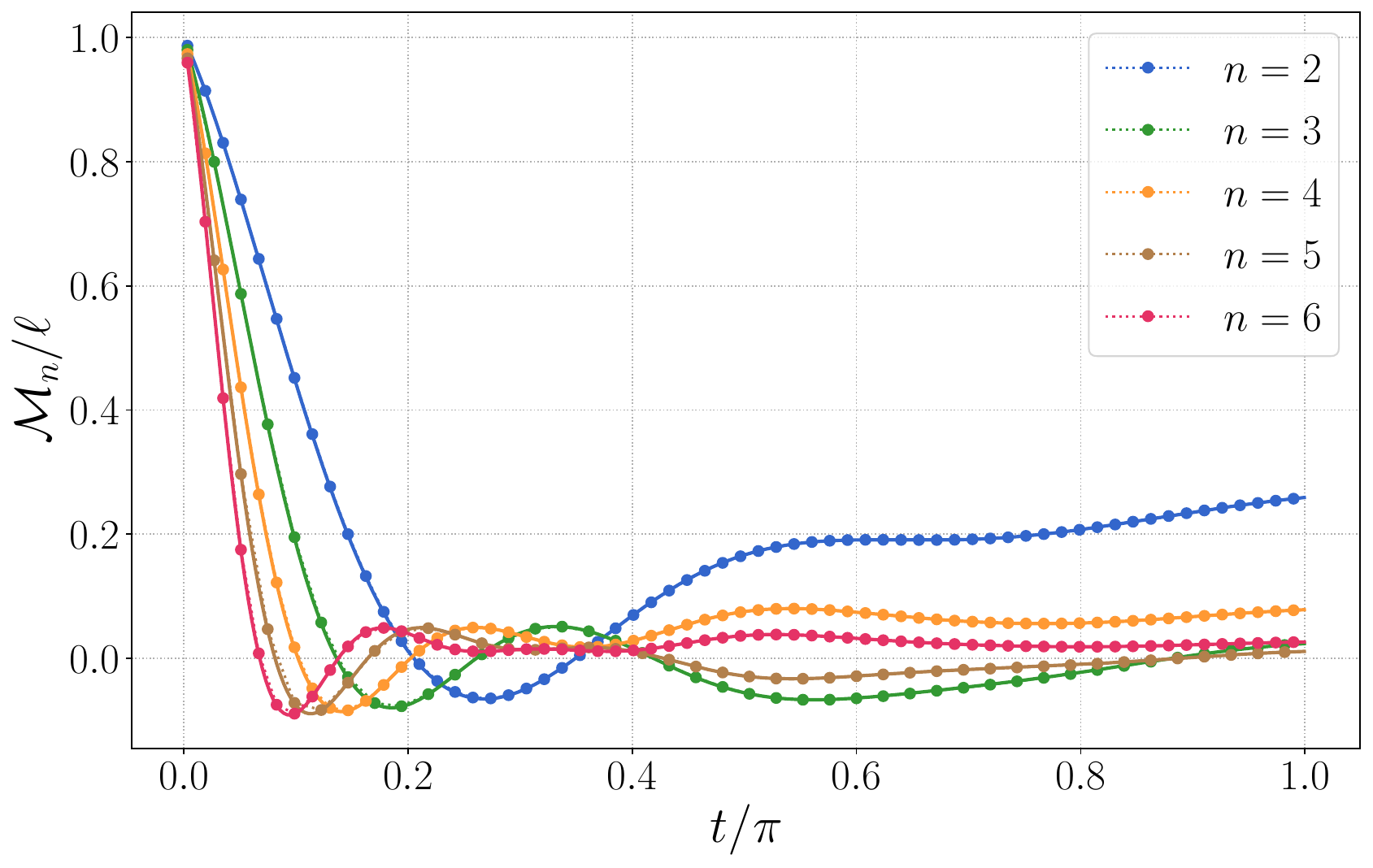}
    \caption{Moments $\M_n$ for various values of $n$ in the dissipative dynamics generated by the tight-binding Hamiltonian from the N\'eel state as a function of $t/\pi$ for $\ell=200$ with gain and loss rates $\gap=0.5$ and $\gam=0.13$. The symbols are the numerical results, and the solid lines are the predictions of Eq.~\eqref{eq:Mn}.  }
    \label{fig:Mn}
    \end{figure}

For arbitrary choices of $\gap$ and $\gam$, the moments $\M_n$ become more cumbersome but can still be computed with techniques discussed in Ref.~\cite{parez2025reduced}. The results involve nontrivial combinations of terms 
\begin{equation}
    T_{p,q} = \int_{-\pi}^{\pi}\frac{\dd k}{2\pi} \eE^{2 p \ir t \cos k}(\ell-\min(2 q|v_k| t,\ell)),
\end{equation}
where $p,q\geqslant 0$ are integers. In particular, $T_{0,0}=\ell$. For $n\leqslant 5$ we have, still in the sub-hydrodynamic regime,
\begin{subequations}\label{eq:Mn}
\begin{equation}
\begin{split}
    \M_2 =& \chi^2 T_{0,0} +b^2 T_{2,1} , \\
     \M_3 =& 3\chi^2 b T_{1,1}+b^3 T_{3,1},\\
     \M_4 =& \chi^4 T_{0,0}+2\chi^2 b^2 T_{0,1}+4\chi^2b^2 T_{2,1}+b^4T_{4,1},\\
     \M_5 =& (5\chi^4b+5\chi^2 b^3)T_{1,1}+5\chi^2 b^3 T_{3,1}+b^5T_{5,1}.
    \end{split}
\end{equation}
For $n\geqslant 6$, there are also contributions from the function $T_{p,q}$ with $q>1$, which signal propagation of entangled quasiparticles with arbitrary large group velocities~$q|v_k|$. These terms give rise to a nested light-cone structure, akin to what is observed in unitary dimer and XY quenches~\cite{parez2025reduced}, as well as in full-counting statistics~\cite{groha2018full}. For instance, for $n=6$ we find
\begin{multline}
     \M_6 =\chi^6 T_{0,0}+6\chi^4b^2 T_{0,1}+(9\chi^4b^2+6\chi^2b^4)T_{2,1} \\ +6\chi^2b^4 T_{4,1}+b^6T_{6,1}+3\chi^2b^4 T_{0,2}.
\end{multline}
\end{subequations}
We verify the analytical predictions of Eq.~\eqref{eq:Mn} against numerical results  in Fig.~\ref{fig:Mn}, and find a perfect agreement. 

It is possible to compute analytically every moment $\M_n$ up to an arbitrary index $n_{\rm max}$ and consider the truncated RLE, $\log f^{\rm trunc}_A(t) = \sum_{n=0}^{n_{\rm max}} c_n \M_n$, similarly as in Ref.~\cite{parez2025reduced} for other unitary quench protocols. However, it is quite challenging to resum the infinite series, and hence it is  not  possible to rigorously investigate nonanalyticities and DQPTs in the dynamics. Since the latter are our only focus here, we do not engage in this procedure and focus solely on numerics for arbitrary gain and loss rates. In the bottom panel of Fig.~\ref{fig:NeelGL}, we observe that gain-only dynamics (identical results hold for the loss-only case) preserves the DQPTs, at the same critical times as in the dissipationless case. Moreover, the introduction of a loss rate $\gam>0$, even extremely small, completely smears out the nonanalyticities, as predicted by our general argument.

\begin{figure}
    \centering
    \includegraphics[width=1\linewidth]{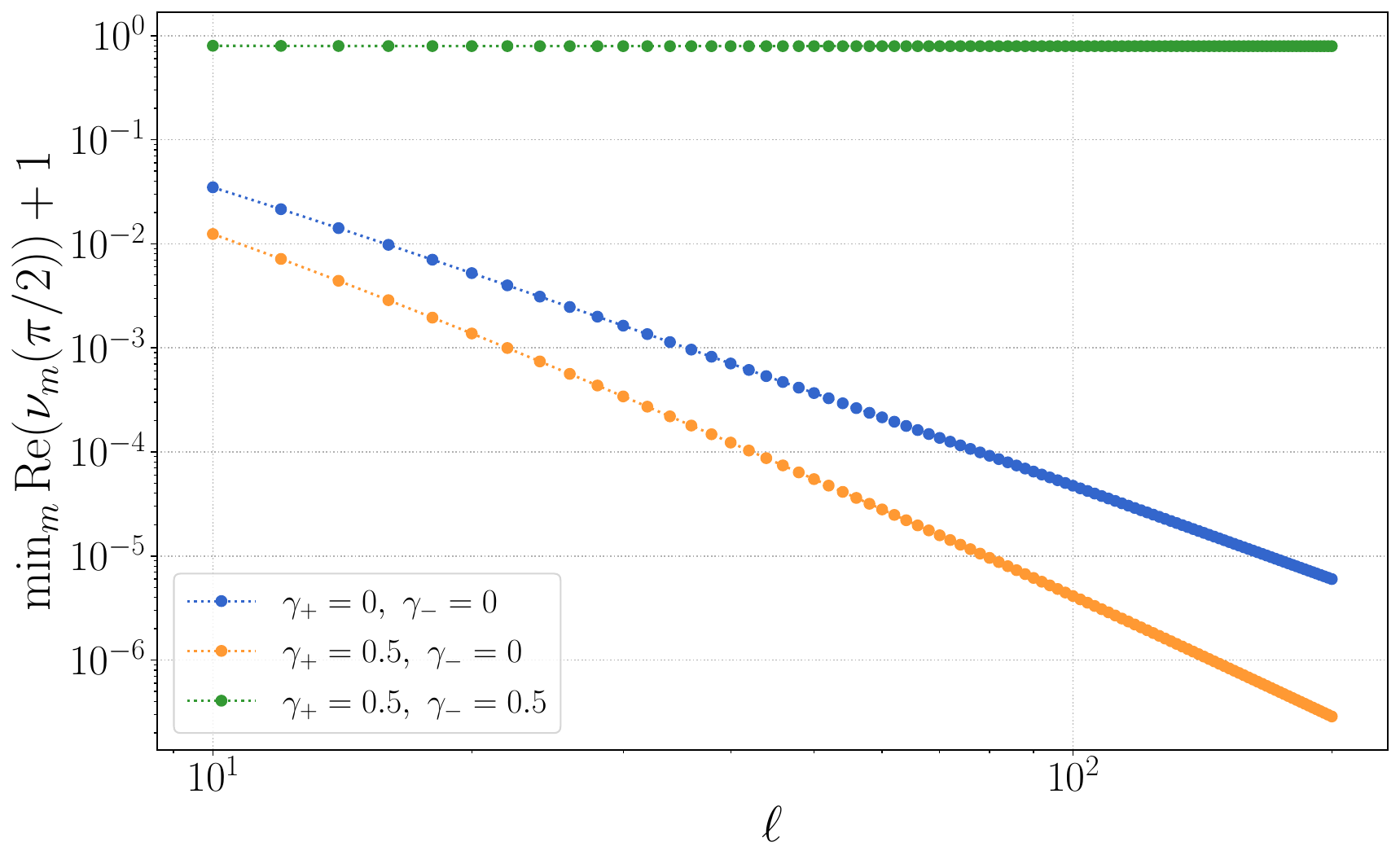}
    \caption{Scaling of the smallest eigenvalue of the matrix $J_0J_A(t)$ as a function of $\ell$ in the N\'eel quench for various values of gain and loss rates $\gamma_\pm$ at the critical time $\pi/2$. }
    \label{fig:lambdaMinNeel}
    \end{figure}

To further verify our results regarding the presence or absence of DQPTs in the N\'eel quench, we investigate the scaling of the eigenvalues $\nu_m(t)$ of the matrix $J_0J_A(t)$ with the system size $\ell$. As discussed in previous sections, a DQPT occur if at least one eigenvalue tends to $\nu_m=-1$ in the large-$\ell$ limit. In Fig.~\ref{fig:lambdaMinNeel}, we numerically investigate the function $\min_m\textrm{Re}(\nu_m(t))$ at the critical time $\tilde t_c^{(0)}~=~\pi/2$ in log-log scale. Both for the unitary ($\gamma_+=\gamma_-=0$) and the gain only ($\gamma_+=0.5$, $\gamma_-=0$) cases, the minimal eigenvalue tends to $-1$, signaling a DQPT. In contrast, for the gain and loss case ($\gamma_+=\gamma_-=0.5$) no eigenvalue scales to $-1$ and hence the RLE is analytical in the thermodynamic limit. This is in perfect agreement with our previous results, and excludes the possibility that the smearing observed in the gain and loss cases might be caused by finite-size effects. 

\section{Results for the dissipative Ising quench}

As a second illustration of our results, we focus on the quantum transverse-field Ising chain. In its fermionic formulation, the model reduces to the Kitaev chain, and the Hamiltonian reads
\begin{equation}
H =  -\frac 12\sum_{j=1}^N\big((c_{j}^\dagger c_{j+1}+c_{j}^\dagger c_{j+1}^\dagger+\textrm{h.c.})+2h c_j^\dagger c_j\big),
\end{equation}
where $h$ is the transverse field. This model does not preserve the fermion number, due to the pairing terms in the Hamiltonian. In the large-$N$ limit, the single-particle spectrum 
reads $\epsilon_k(h) = \sqrt{h^2+1-2h\cos k}$ for $k \in [-\pi, \pi]$. The diagonalization of the model requires the introduction of Majorana operators, defined as $a_{2m-1} = c_m +c_m^\dagger$ and $a_{2m} = \ir (c_m -c_m^\dagger)$. They satisfy the anticommutation relation $\{a_j, a_k\} = 2\delta_{j,k}$. We consider a quench protocol where the initial state $|\psi_0\rangle$ is the ground state of the Ising chain with transverse field $h_0$, and for $t>0$ the time evolution is governed by the model with transverse field $h\neq h_0$, and with gain and loss rates $\gamma_\pm$.

In situations where the fermion number is not preserved, the covariance $\ir \Gamma_A(t)$ is defined as
\begin{equation}
\label{eq:cov}
\ir[\Gamma_A(t)]_{m,n}=\begin{pmatrix}\langle a_{2m-1}a_{2n-1} \rangle-\delta_{m,n}& \langle  a_{2m-1}a_{2n}\rangle  \\ \langle a_{2m} a_{2n-1}\rangle &  \langle  a_{2m} a_{2n}\rangle-\delta_{m,n}
\end{pmatrix} ,
\end{equation}
for $m,n=1,2,\dots,\ell$, where $\langle a_x a_y\rangle = \Tr(\rho_A(t) a_xa_y)$ is the time-dependent two-point correlation function. Hence, the dimension of $\Gamma_A$ is $2\ell$. The (unnormalized) RLE reads in this case \cite{parez2022symmetry,parez2025reduced} 
\begin{equation}\label{eq:fdet2}
    f_A(t) = \sqrt{\det\left(\frac{\mathbb{1}_{2\ell}-\Gamma_0\Gamma_A(t)}{2}\right)},
\end{equation}
where $\Gamma_0\equiv\Gamma_A(0)$, and the logarithmic unnormalized RLE is $\zeta_A(t)=-1/\ell \log f_A(t)$. For the dissipationless quench, the time-dependent correlation functions are known analytically \cite{calabrese2005evolution,fc-08}. We have
\begin{equation}
\label{eq:Toeplitz}
\begin{split}
 [\tilde \Gamma_A(t)]_{m,n}&=\int_{-\pi}^\pi \frac{\dd k}{2\pi}\eE^{-ik(m-n)} \tilde \Gamma_k(t) ,\\ 
\tilde \Gamma_k(t) &= \begin{pmatrix} -F_k(t) & G_k(t)\\
-G_{-k}(t) & F_k(t)\end{pmatrix} ,
\end{split}
\end{equation}
where 
\begin{equation}
\label{eq:FG}
\left\{
\begin{aligned}
  F_k(t)&= \ir \sin \Delta_k \sin (2 \epsilon_k(h) t),  \\
  G_k(t)&= \eE^{- \ir \theta_k} \left(\cos \Delta_k+\ir \sin \Delta_k \cos (2 \epsilon_k(h) t) \right).
\end{aligned}
\right. 
\end{equation}
Here, $\theta_k$ is the Bogoliubov angle, and $\Delta_k=\theta_k-\theta_k^{(0)}$ is the difference between the angles after and before the quench. They read
\begin{equation}
\label{eq:quenchparameters}
\begin{split}
    \eE^{\ir \theta_k}&=\frac{\cos k-h+\ir  \sin k}{\epsilon_k(h)},\\
    \cos \Delta_k &=\frac{1+hh_0-(h+h_0)\cos k}{\epsilon_k(h)\epsilon_{k}(h_0)}.\\
\end{split}
\end{equation}
This unitary dynamics exhibits temporal nonanalyticities signaling DQPTs at times $\tilde t_c^{(m)}=(m+1/2)t^*$, where $m\geqslant 0$ is an integer, and $t^*$ is~\cite{heyl2013dynamical,parez2025reduced}
\begin{equation}
    t^*=\frac{\pi}{\sqrt{\Big(h-\frac{1+h_0 h}{h_0+h}\Big)^2+1-\Big(\frac{1+h_0 h}{h_0+h}\Big)^2}}.
\end{equation}

In the presence of gain and loss dissipation with rates~$\gamma_\pm$, the covariance matrix reads~\cite{alba2022hydrodynamics,caceffo2024fate}
\begin{equation}
\label{eq:ToeplitzDissip}
\begin{split}
[&\Gamma_A(t)]_{m,n}=\int_{-\pi}^\pi \frac{\dd k}{2\pi}\eE^{-ik(m-n)}\Gamma_k(t) ,\\ 
&\Gamma_k(t) = b(t) \tilde \Gamma_k(t) + \chi(t) \begin{pmatrix} 0& \cos\theta_k \eE^{\ir \theta_k}\\
-\cos\theta_k \eE^{-\ir \theta_k} &0\end{pmatrix} ,
\end{split}
\end{equation}
where $b(t),\chi(t)$ are defined in Eqs.~\eqref{eq:b} and \eqref{eq:chi}, and $\tilde \Gamma_k(t)$ is given in Eq.~\eqref{eq:Toeplitz}. Let us recall that we use the tilde notation to refer to quantities in the dissipationless dynamics. 

From Eqs.~\eqref{eq:fdet2} and \eqref{eq:ToeplitzDissip}, we can generalize the discussion relating DQPTs in dissipative dynamics with eigenvalues of covariance matrices to the case where the fermion number is not conserved. Using the same arguments as in the fermion-number-preserving case, we conclude that DQPTs occur at times $t_c$ such that the product of covariance matrices $\ir \Gamma_0 \ir \Gamma_A(t_c)$ has at least one eigenvalue approaching $-1$ in the large-$\ell$ limit. Moreover, such DQPTs are ruled out in the presence of both gain and loss processes. Dissipative DQPTs can thus only potentially occur for gain-only or loss-only dynamics, at the critical times of the unitary dynamics, $t_c=\tilde t_c$. 

Similarly as before, to analytically investigate the RLE, one needs to (i) decompose the logarithmic RLE in terms of its moments, $\log f_A(t) = 1/2\sum_{n=0}^\infty c_n \M_n$, defined here as $\M_n = (-1)^n\Tr((\Gamma_0\Gamma_A(t))^n)$, (ii) find an exact expression for each moment, and (iii) resum the series. However, this task already proves to be extremely challenging in the dissipationless case, because the moments $\widetilde \M_n$ exhibit a nested lightcone structure \cite{parez2025reduced}. As we discussed for the dissipative N\'eel quench, the presence of dissipation does not simplify this lightcone structure. On the contrary, dissipation enhances it, such that the structure emerges even in cases where it is not present in the unitary~dynamics.

\begin{figure}
    \centering
    \includegraphics[width=1\linewidth]{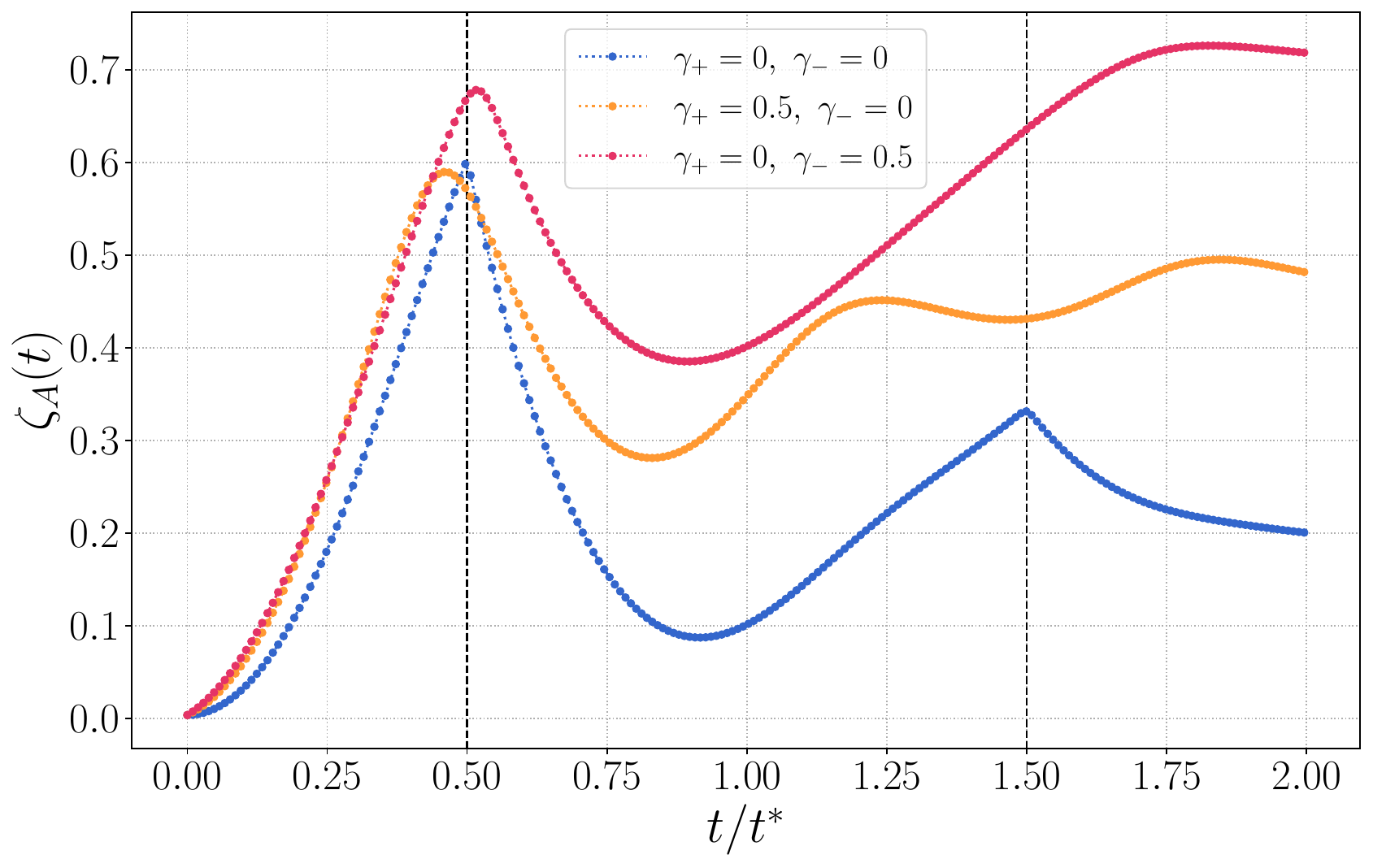}\\
    \includegraphics[width=1\linewidth]{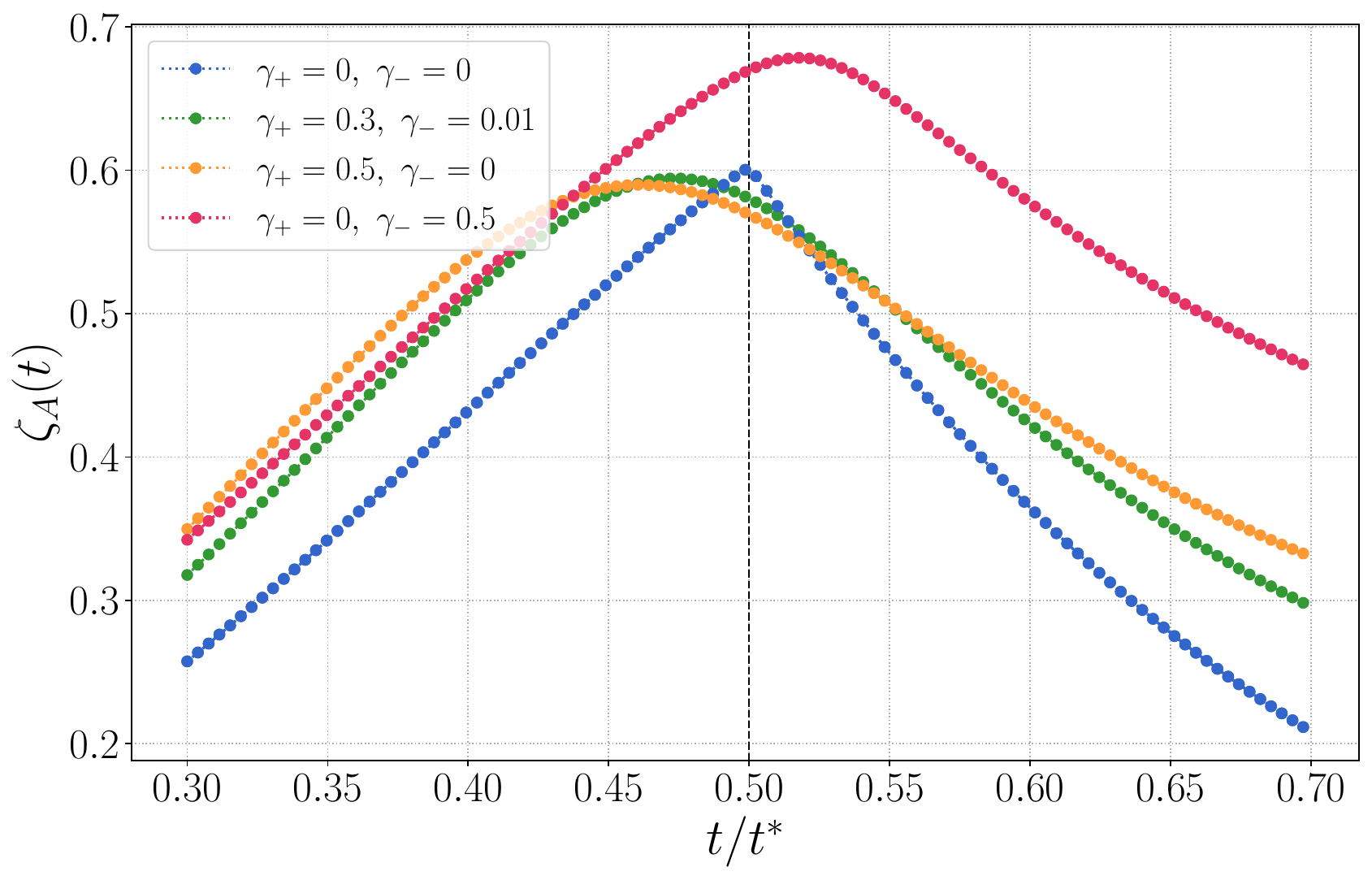}
    \caption{Logarithmic RLE $\zeta_A(t)$ in the dissipative Ising quench with $\ell=200$, $h_0=0$ and $h=2.6$ as a function of $t/t^*$ for various values of gain and loss rates. The symbols are obtained by numerical diagonalization of the covariance matrices from Eqs.~\eqref{eq:fdet2} and \eqref{eq:ToeplitzDissip}, and the vertical dashed lines indicate the position of the DQPTs in the dissipationless dynamics. The lower panel is a zoom on the first critical time, to highlight the smearing of the DQPT from the dissipationless dynamics, even in the gain-only and loss-only cases. }
    \label{fig:IsingDissip}
    \end{figure}

For the dissipative Ising quench, we thus investigate the dynamics of the logarithmic RLE numerically, and report our results in Fig.~\ref{fig:IsingDissip}. We observe that, in contrast with the dissipative N\'eel quench, the DQPTs of the unitary dynamics are smeared out, even for gain-only and loss-only processes. This is not in contradiction with our general results, since we stress again that the presence of DQPTs in the unitary dynamics is a necessary but not sufficient condition for DQPTs in the dissipative case. Interestingly, the first peak at $t=t^*/2$ is still visible (but smeared out) in the gain-only and loss-only dynamics, but the second one at $t=3t^*/2$ is not detectable in the dissipative dynamics. In the lower panel of Fig.~\ref{fig:IsingDissip}, we also observe that the presence of both gain and loss processes, even for small values of $\gamma_\pm$, completely smears out the nonanalyticities, as predicted by our general result.

\begin{figure}
    \centering
    \includegraphics[width=1\linewidth]{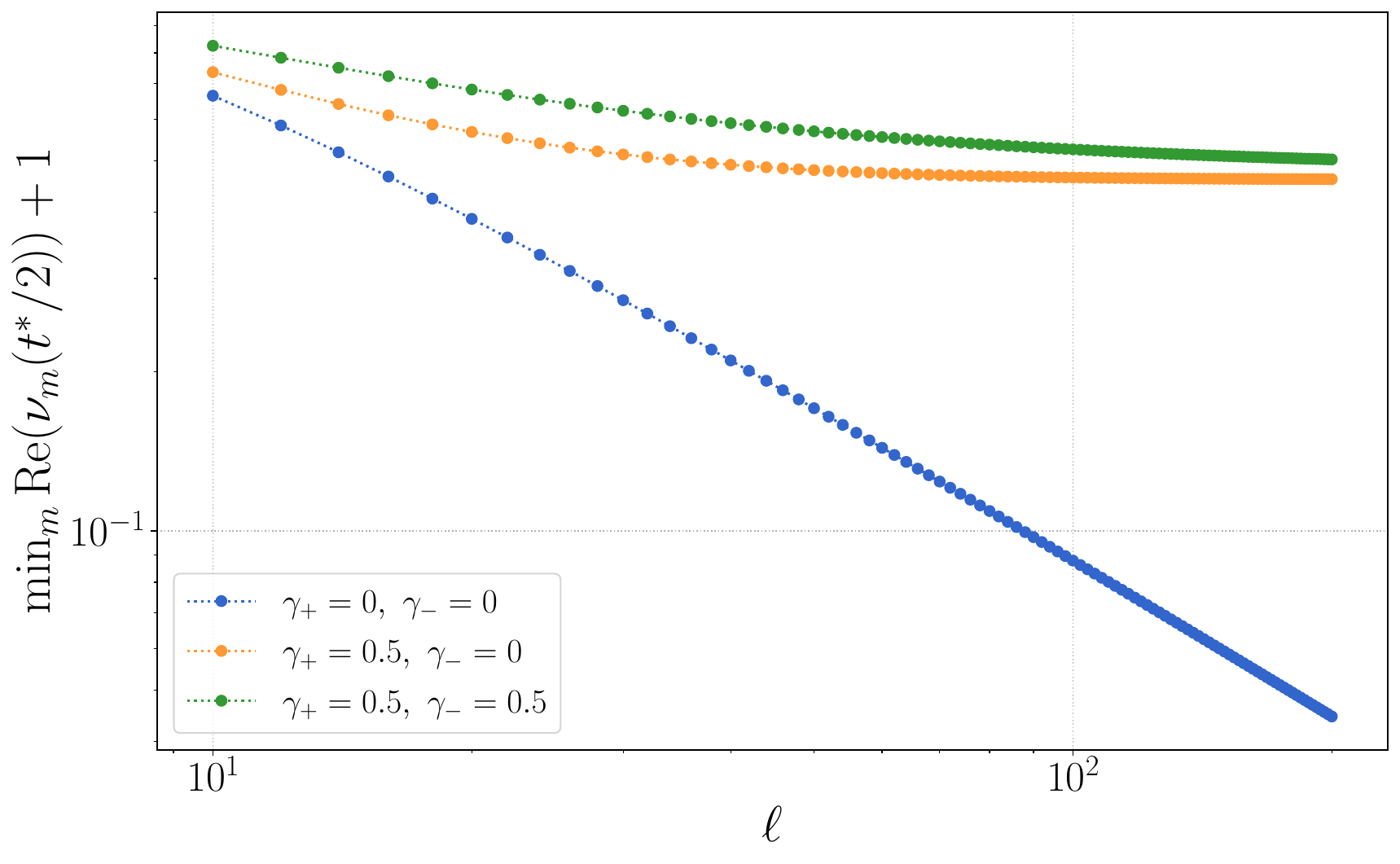}
    \caption{Scaling of the smallest eigenvalue of the matrix~$(-\Gamma_0\Gamma_A(t))$ as a function of $\ell$ in the Ising quench for $h_0=0$, $h=2.6$ and various values of gain and loss rates $\gamma_\pm$ at the critical time $t^*/2$. }
    \label{fig:lambdaMinIsing}
    \end{figure}

Similarly to the N\'eel quench, we further verify our results for the Ising quench by investigating the scaling of the eigenvalues $\nu_m$ of the matrix $(-\Gamma_0\Gamma_A(t))$ with the system size $\ell$.  In Fig.~\ref{fig:lambdaMinIsing}, we numerically investigate the function $\min_m\textrm{Re}(\nu_m(t))$ at the critical time $\tilde t_c^{(0)}~=~t^*/2$ in log-log scale. For the unitary case ($\gamma_+=\gamma_-=0$), the minimal eigenvalue approaches $-1$ in the large-$\ell$ limit, signaling a DQPT. For the gain only ($\gamma_+=0.5$, $\gamma_-=0$) and gain and loss cases ($\gamma_+=\gamma_-=0.5$), no eigenvalue scales to $-1$ and hence there is no DQPT. This is in perfect agreement with our previous numerical results for the Ising quench. This analysis excludes the possibility that the smearing observed in the gain and gain and loss cases might be caused by finite-size effects.

\section{Conclusions}
We studied the fate of dynamical quantum phase transitions in dissipative free-fermionic systems using the reduced Loschmidt echo as a probe. Our results establish a clear link between unitary and dissipative dynamics: dynamical quantum phase transitions in the open system can only occur if they are already present in the corresponding unitary evolution, and only in the cases of pure gain or pure loss. By contrast, the simultaneous presence of both processes—no matter how small—completely smears out the nonanalytic behavior. We confirmed these findings analytically and numerically for the quench from the N\'eel state in the tight-binding chain, as well as for the quantum Ising chain, both with gain and/or loss dissipation. Moreover, we showed that the presence of dissipation can generate a nested lightcone structure in the dynamics of the reduced Loschmidt echo, even if this structure is not present in the corresponding unitary evolution. These findings highlight the reduced Loschmidt echo as a versatile tool to investigate nonequilibrium singularities in both isolated and open settings, and provide a rigorous framework for future experimental realizations. In particular, it would be extremely interesting to verify if the smearing (or lack thereof) of dynamical quantum phase transitions, as well as the nested lightcone structure, can be observed in cold-atom experiments which allow for gain and loss~dissipation.

\textbf{Acknowledgments.---}GP thanks Cl\'ement Berthiere for useful discussions. This study was carried out within the National Centre on HPC, Big Data and Quantum Computing - SPOKE 10 (Quantum Computing) and received funding from the European Union NextGenerationEU - National Recovery and Resilience Plan (NRRP) – MISSION 4 COMPONENT 2, INVESTMENT N. 1.4 – CUP N. I53C22000690001. This work has been supported by the project ``Artificially devised many-body quantum dynamics in low dimensions - ManyQLowD'' funded by the MIUR Progetti di Ricerca di Rilevante Interesse Nazionale (PRIN) Bando 2022 - grant 2022R35ZBF.
%

\providecommand{\href}[2]{#2}\begingroup\raggedright\endgroup

\end{document}